# Applications of Quantum Computing for Investigations of Electronic Transitions in Phenylsulfonyl-carbazole TADF Emitters

Qi Gao[1,2*], Gavin O. Jones[3*], Mario Motta[3], Michihiko Sugawara[2], Hiroshi C. Watanabe[2], Takao Kobayashi[1], Eriko Watanabe[1,2], Yu-ya Ohnishi[2,4], Hajime Nakamura[2,5], Naoki Yamamoto[2]

[1]Mitsubishi Chemical Corporation, Science & Innovation Center,
1000, Kamoshida-cho, Aoba-ku, Yokohama 227-8502, Japan

[2]Quantum Computing Center, Keio University, Hiyoshi 3-14-1, Kohoku, Yokohama 223-8522, Japan

[3]IBM Quantum, IBM Research – Almaden, 650 Harry Road, San Jose, CA 95120, USA

[4]Materials Informatics Initiative, Yokkaichi Research Center,
JSR Corporation, 100 Kawajiri-cho, Yokkaichi, Mie, 510-8552, Japan

[5]IBM Research – Tokyo, 19-21 Nihonbashi Chuo-ku, Tokyo 103-8510, Japan

[6]Department of Applied Physics and Physico-Informatics, Keio University, Hiyoshi 3-14-1, Kohoku, Yokohama 223-8522, Japan

A quantum chemistry study of the first singlet ($S_1$) and triplet ($T_1$) excited states of phenylsulfonyl-carbazole compounds, proposed as useful thermally activated delayed fluorescence (TADF) emitters for organic light emitting diode (OLED) applications, was performed with the quantum Equation-Of-Motion Variational Quantum Eigensolver (qEOM-VQE) and Variational Quantum Deflation (VQD) algorithms on quantum simulators and devices. These quantum simulations were performed with double zeta quality basis sets on an active space comprising the highest occupied and lowest unoccupied molecular orbitals (HOMO, LUMO) of the TADF molecules. The differences in energy separations between $S_1$ and $T_1$ ($\Delta E_{st}$) predicted by calculations on quantum simulators were found to be in excellent agreement with experimental data. Differences of 16 and 88 *mHa* with respect to exact energies were found for excited states by using the qEOM-VQE and VQD algorithms, respectively, to perform simulations on quantum devices without error mitigation. By utilizing error mitigation by state tomography to purify the quantum states and correct energy values, the large errors found for unmitigated results could be improved to differences of, at most, 3 *mHa* with respect to exact values. Consequently, excellent agreement could be found between values of $\Delta E_{st}$ predicted by quantum simulations and those found in experiments.

## Introduction

One of the major constraints of modern electronic structure methods used for quantum chemical calculations on classical computing architecture is the difficulty of finding eigenvalues of eigenvectors of the electronic Hamiltonian. [1] The advent of quantum computing, which has demonstrated tremendous synergistic advances in both hardware and software capabilities in recent years, may provide invaluable support in the investigation of the electronic structure of molecules and materials, especially with regards to dynamical properties. However, quantum computing is still, in many ways, a nascent technology, and quantum devices are currently hamstrung by noisiness and short decoherence times. Thus, quantum algorithms, such as the Variational Quantum Eigensolver (VQE) algorithm, [2] are currently being used to find eigenvalues for approximate Ansätze suitable for constructing relatively short circuits that can be used for quantum chemistry calculations. [3]

---

[*]Corresponding authors. E-mail: caoch@user.keio.ac.jp; gojones@us.ibm.com

A number of quantum computing use cases for chemistry have been investigated on quantum devices in the recent past. One of the first applications highlighted the construction of ground state dissociation profiles of hydrogen, lithium hydride and beryllium hydride via computation using an IBM Q quantum device. [3] Researchers have also performed computational studies on the mechanism of the rearrangement of the lithium superoxide dimer using larger devices. [4] In the latter case, the orbital determinants of the stationary points were extensively examined to identify a suitable orbital active space for which a reduced number of qubits could be used for computation given the current limitations of these devices. Simulations performed on classical hardware reproduced calculations involving the exact eigensolver known as Full Configuration Interaction (Full CI or FCI) for this process. However, calculations performed on quantum devices using hardware efficient Ansätze such as *Ry* were less capable of reproducing exact energies due to noise.

In addition to using the VQE algorithm to compute energies for the dissociation profiles of molecules and the potential energy surfaces for reactions, various algorithms, among them the Quantum Equation-of-Motion VQE (qEOM-VQE) [5] and Variational Quantum Deflation (VQD) algorithms [6], have been developed in order to compute excited states of molecules. In particular, Ollitrault et al. have demonstrated computation of the first three excited states on the dissociation profile for lithium hydride on IBM Quantum hardware. [5]

Here, we take an initial step towards the application of algorithms such as qEOM-VQE and VQD to determine excitation energies of industrially relevant molecules. The subject of this study is thermally activated delayed fluorescence (TADF) emitters suitable for organic light emitting diode (OLED) applications. [7] Once the separation between the first singlet ($S_1$) and triplet ($T_1$) excited states ($\Delta E_{st}$) is sufficiently small, non-emissive $T_1$ excitons can be thermally excited to an emissive S1 state, providing the emission mechanism of TADF emitters. The mechanism enables OLED devices comprised of TADF emitters to potentially perform with 100% internal quantum efficiency, in contrast to devices comprising conventional fluorophores for which the quantum efficiencies are inherently limited to 25%. TADF emitters have been demonstrated as next-generation emission materials for a range of fluorophores and have been utilized in the fabrication of efficient OLEDs. [7–12]

To date there have been a number of computational studies on predictions of $\Delta E_{st}$ to aid in the design of novel TADF materials. [12–17] Based on those investigations, a well-established strategy for designing TADF molecules involves identifying emitters in which HOMO and LUMO are spatially separated and localized on appropriate moieties. Previous demonstrations have shown that a series of phenylsulfonyl-9H-carbazole (PSPCz) molecules are useful as TADF emitters due to the fact that the $\Delta E_{st}$ of these molecules can be finely tuned by modifying their electronic properties. [18] We have elected to perform quantum chemical investigations on the molecules, PSPCz, 2F-PSPCz, 4F-PSPCz (Figure 1), using quantum devices and simulators.

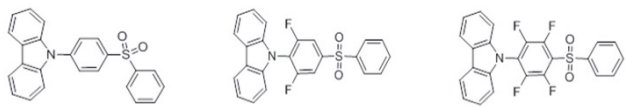

**Fig 1**. Molecular structures of the PSPCz, 2F-PSPCz and 4F-PSPCz molecules.

Notably, all three molecules of interest require the use of many more qubits than are currently available on useful quantum hardware or that can be reliably simulated using classical architecture. Thus, qubit reduction techniques must be applied in order to simulate the transition amplitudes of interest to this study. We have reduced the number of spatial orbitals to those that are absolutely necessary to describe the processes under investigation and have thus focused on transitions involving the HOMO and LUMO active space for each molecule. This strategy has allowed reduction of the number of qubits to just two after applying spin parity reduction.

Simulations of the qEOM-VQE and VQD algorithms were performed on classical devices using heuristic Ansätze, in order to assess the accuracy of these techniques. Particular attention was paid to comparisons with experimental data [18] which indicate that these procedures produce meaningful data for predicting the $\Delta E_{st}$ of TADF emitters. Moreover, we have shown that these simulations can provide a detailed picture about how structural variation in phenylsulfonyl/carbazole molecules tunes the $S_1$ and $T_1$ excitations. Finally, we have found that simulations performed on a quantum device without error mitigation were much less reliable than results provided by quantum simulators. This is attributed to the fact that the quantum state of the ground state predicted on devices is a mixed state.

We have examined various strategies to solve this issue and have found that the most accurate procedure involves the use of state tomography to purify the mixed ground state obtained from the quantum device prior to application of readout error mitigation to the calculation involving the excited state. We note that since the state tomography purification approach can, in principle, be applied to other methods for computing excited states on quantum computers [19,20], and can be extended to systems requiring large number of qubits by purifying the quantum state using a computationally inexpensive iterative approach, [21] or by applying quantum principal component analysis, [22] or variational quantum state diagonalization, [23] we believe that it can be generally useful for quantum chemistry simulations of excited states on near-term quantum devices.

**Methods**

Four computational steps were performed in order to evaluate the electronic transitions between the singlet and triplet excited states of the PSPCz molecules: (i) structural optimization on classical architecture, (ii) computation of ground state (GS) energies on quantum simulators and quantum devices utilizing the HOMO-LUMO active space, (iii) error mitigation applied to ground state energies obtained from quantum devices, (iv) computation of the first singlet and triplet excited states ($S_1$ and $T_1$), using quantum simulators and quantum devices.

Optimized geometries of the first triplet excited state ($T_1$) of all PSPCz molecules were generated using the TDDFT routines contained in the Gaussian16 computational suite of programs [24] using the CAM-B3LYP/6-31G(d) method and basis set. The Tamm-Dancoff approximation (TDA) [25], which has been reported to be accurate for the calculation of triplet excited states [26], was employed for all TDDFT calculations. For the sake of simplicity, the optimized geometries of the $T_1$ excited states were used for both the $S_1$ and $T_1$ excited states based on the fact that the HOMO-LUMO overlap (i.e. exchange integral) in TADF molecules is so small that the properties of the $S_1$ and $T_1$ states should be quite similar, except for the spin component, and the structures of these excited states should also be similar.

The HOMO and LUMO was chosen as the active space for calculating the ground state ($S_0$), and $S_1$ and $T_1$ excited states because the HOMO and LUMO of the target molecules are spatially fairly well-separated as shown in Figure 2, and both the $S_1$ and $T_1$ states can be well described by a transition arising from the HOMO, with atomic orbitals localized on the carbazole moiety, to the LUMO, with atomic orbitals localized on the diarylsulfone ($ArSO_2Ar$-) moieties.

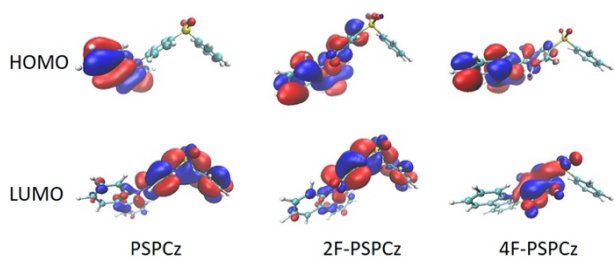

**Fig 2**. HOMO and LUMO orbitals of the triplet state optimized structures of PSPCz, 2F-PSPCz and 4F-PSPCz, respectively.

Calculations were performed to obtain ground state energies on quantum simulators and devices by utilizing the VQE algorithm and the *Ry* heuristic Ansätze with a circuit depth equal to 1. Energies were computed using the STO-3G minimal basis set and the split-level double zeta basis set 6-31G(d). The parity mapping technique [27], which allows the number of qubits to be reduced by two, was used to map molecular spin-orbitals to qubits. The Aqua module contained in Qiskit version 0.14 [28] with an interface to the PySCF [29] program was used for all VQE calculations.

Calculations were performed using the `statevector` and `qasm` simulators contained in the Aer module of Qiskit, as well as on the `ibmq_boeblingen` and `ibmq_singapore` 20-qubit quantum devices. The `statevector` simulator computes wavefunctions and expectation values exactly, by means of linear algebra operations. In contrast, the `qasm` simulator samples quantum mechanical probability distributions that can be determined exactly by means of linear algebra operations, or by approximating actual decoherence processes occurring on hardware when used with noise models. The user specifies the number of statistical samples, or shots, drawn from the target probability distribution when using the `qasm` simulator.

The Sequential Least SQuares Programming (SLSQP) method [30], which uses the exact gradient for energy minimization and has been found to provide accurate predictions without noise present, [31] was used for calculations using the `statevector` simulator. The Simultaneous Perturbation Stochastic Approximation (SPSA) [32,33] method, which is a good optimizer in the presence of noise, [34] was used for calculations on quantum devices.

The readout error mitigation and quantum state tomography techniques provided in Qiskit were used to mitigate errors in the generation of variational parameters, $\theta$, and ground state energies.

The four basis states of the two-qubit system, namely |00>, |01>, |10> and |11>, were measured for the readout error mitigation technique. A vector with the probability for the four basis states was created for every measurement, and the combination of all four vectors resulted in the formation of a 4x4 matrix from which a calibration matrix was generated by least-squares fitting. The calibration matrix was applied to the measurement of Pauli terms in the VQE calculation to correct the energy value of ground state. The values of $\theta$ in *Ry* were then calculated using a classical optimizer..

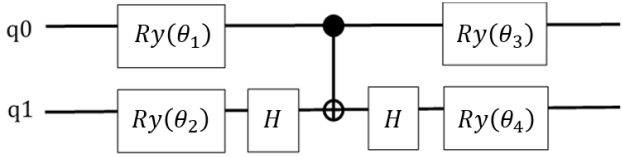

**Fig 3**. Quantum circuit implementing the *Ry* Ansätze for 2 qubits.

$$\cos\left(\frac{\theta_1}{2}\right)\cos\left(\frac{\theta_2}{2}\right)\cos\left(\frac{\theta_3}{2}\right)\cos\left(\frac{\theta_4}{2}\right) - \cos\left(\frac{\theta_2}{2}\right)\cos\left(\frac{\theta_4}{2}\right)\sin\left(\frac{\theta_1}{2}\right)\sin\left(\frac{\theta_3}{2}\right)$$
$$-\cos\left(\frac{\theta_1}{2}\right)\cos\left(\frac{\theta_3}{2}\right)\sin\left(\frac{\theta_2}{2}\right)\sin\left(\frac{\theta_4}{2}\right) - \sin\left(\frac{\theta_1}{2}\right)\sin\left(\frac{\theta_2}{2}\right)\sin\left(\frac{\theta_3}{2}\right)\sin\left(\frac{\theta_4}{2}\right) \quad (1)$$

$$\cos\left(\frac{\theta_2}{2}\right)\cos\left(\frac{\theta_3}{2}\right)\cos\left(\frac{\theta_4}{2}\right)\sin\left(\frac{\theta_1}{2}\right) + \cos\left(\frac{\theta_1}{2}\right)\cos\left(\frac{\theta_2}{2}\right)\cos\left(\frac{\theta_4}{2}\right)\sin\left(\frac{\theta_3}{2}\right)$$
$$+\cos\left(\frac{\theta_3}{2}\right)\sin\left(\frac{\theta_1}{2}\right)\sin\left(\frac{\theta_2}{2}\right)\sin\left(\frac{\theta_4}{2}\right) - \cos\left(\frac{\theta_1}{2}\right)\sin\left(\frac{\theta_2}{2}\right)\sin\left(\frac{\theta_3}{2}\right)\sin\left(\frac{\theta_4}{2}\right) \quad (2)$$

$$\cos\left(\frac{\theta_1}{2}\right)\cos\left(\frac{\theta_3}{2}\right)\cos\left(\frac{\theta_4}{2}\right)\sin\left(\frac{\theta_2}{2}\right) + \cos\left(\frac{\theta_4}{2}\right)\sin\left(\frac{\theta_1}{2}\right)\sin\left(\frac{\theta_2}{2}\right)\sin\left(\frac{\theta_3}{2}\right)$$
$$+\cos\left(\frac{\theta_1}{2}\right)\cos\left(\frac{\theta_2}{2}\right)\cos\left(\frac{\theta_3}{2}\right)\sin\left(\frac{\theta_4}{2}\right) - \cos\left(\frac{\theta_2}{2}\right)\sin\left(\frac{\theta_1}{2}\right)\sin\left(\frac{\theta_3}{2}\right)\sin\left(\frac{\theta_4}{2}\right) \quad (3)$$

$$-\cos\left(\frac{\theta_3}{2}\right)\cos\left(\frac{\theta_4}{2}\right)\sin\left(\frac{\theta_1}{2}\right)\sin\left(\frac{\theta_2}{2}\right) + \cos\left(\frac{\theta_1}{2}\right)\cos\left(\frac{\theta_4}{2}\right)\sin\left(\frac{\theta_2}{2}\right)\sin\left(\frac{\theta_3}{2}\right)$$
$$+\cos\left(\frac{\theta_2}{2}\right)\cos\left(\frac{\theta_3}{2}\right)\sin\left(\frac{\theta_1}{2}\right)\sin\left(\frac{\theta_4}{2}\right) + \cos\left(\frac{\theta_1}{2}\right)\cos\left(\frac{\theta_2}{2}\right)\sin\left(\frac{\theta_3}{2}\right)\sin\left(\frac{\theta_4}{2}\right) \quad (4)$$

The quantum state tomography technique involves purification of the state by diagonalizing the density matrix and selection of the pure state with the maximum eigenvalue resulting from this process. The ground state energy is then refined by computing the expectation value of the Hamiltonian with respect to the obtained pure state. The *Ry θ* values corresponding to the purified ground state were determined by fitting the vector of the Ansätze to the purified maximum eigenstate. Figure 3 shows that the *Ry* circuit for the two-qubit system produces an output state Y(*q*) that can be represented as a linear combination of the four basis states |00>, |01>, |10> and |11>, with coefficients given by Equation 1.

Note that 300 iterations were performed without error mitigation prior to application of the readout error mitigation or quantum state tomography techniques to obtain a quantum state which is close to the ground state generated without noise. The optimized parameters obtained from VQE without error mitigation was used as the initial state for VQE calculations combined with readout error mitigation, and the optimized wavefunction was determined by using previously published protocols. [34] The optimized wavefunction obtained from the use of quantum state tomography was determined by using the density matrix of the final result obtained from VQE without error mitigation contained in the `tomography` module of the Ignis element of Qiskit.

The optimized wavefunction found by application of the readout error mitigation and quantum state tomography techniques to ground states computed with VQE were used to compute the energies of the $S_1$ and $T_1$ excited states by the use of the qEOM-VQE [5] and VQD [6,31] algorithms. The values of matrix elements of the EOM (equation of motion) were computed on quantum devices using the qEOM-VQE algorithm with the GS obtained from VQE as the reference state. The matrix was then diagonalized on a classical computer and the first and second eigenvalues regarded as the energies of the $T_1$ and $S_1$ states relative to the $S_0$ ground state. qEOM-VQE calculations were performed using the `statevector` simulator and on the `ibmq_boeblingen` quantum device.

The VQD algorithm is constructed by defining a new Hamiltonian based on the original molecular Hamiltonian and the overlap formed by the wavefunctions of the reference state and the excited states ($T_1$ or $S_1$). [6] This new Hamiltonian was then used by VQD to compute the energies of the $T_1$ or $S_1$ excited states. VQD calculations were performed using the `statevector` simulator and on the `ibmq_singapore` quantum device. For VQD calculations using the `statevector` simulator, the $T_1$ and $S_1$ states were specified by adding the expectation value of the total spin ($\langle S^2 \rangle$) as a penalty term to the cost function [31] and the ground state obtained from VQE used as both the initial state and the reference state. However, this penalty term was not used for VQD calculations on quantum devices due to the effect of noise on the accuracy of predicted expectation values. Instead, the GS obtained from VQE was used as the initial reference state as usual, to perform a VQD calculation to obtain the first excited state, $T_1$, which was in turn used as the initial state, and then both $S_0$ and $T_1$ were used as reference states in another VQD calculation to obtain the second excited state, $S_1$.

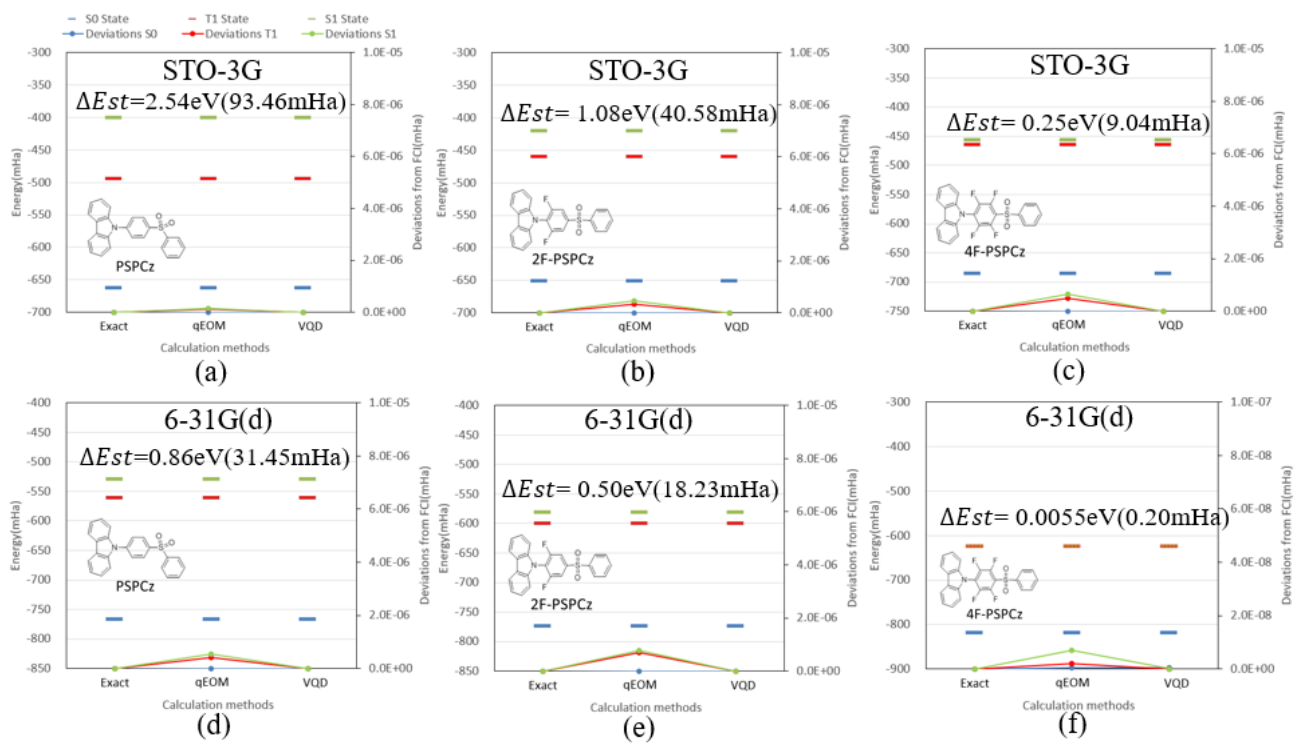

**Fig 4.** Energies of the S0, T1 and S1 (represented in blue, red and green) states for (a,d) PSPCz, (b,e) 2F-PSPCz and (c,f) 4F-PSPCz from exact diagonalization, qEOM and VQD calculations with the statevector simulator using the (a-c) STO-3G, and (d-f) 6-31G(d) basis sets.

## Results and Discussion
### A. Calculations with the `statevector` simulator

The energies of the $S_0$, $S_1$ and $T_1$ states calculated by the Full CI, and *Ry* methods using qEOM-VQE and VQD algorithms with the STO-3G and 6-31G(d) basis sets on the `statevector` simulator are shown in Figure 4. Similar results are predicted by the use of *Ry* with the qEOM-VQE and VQD algorithms for the energies of the $S_0$ state due to the fact that both algorithms use the VQE method to the zeroth approximation for the ground state.

Although qEOM-VQE and VQD use different strategies to calculate excited states, both predict very similar energies for the $S_1$ and $T_1$ states. Both qEOM-VQE and VQD calculations predict energies that reproduce those predicted by Full CI. These results imply that both the qEOM-VQE and VQD algorithms can accurately predict of excited state energies of TADF emitters.

As suggested by the data in Figure 4, predicted excited state energies are basis set dependent. The $S_1$ state is increasingly lowered by increasing the electron-deficiency of the phenylcarbazole by increasing the number of fluorine substituents on going from PSPCz to 2F-PSPCz and then to 4F-PSPCz for the STO-3G basis set. Interestingly, while the energy of the $T_1$ state increases by changing from PSPCz to the more electron-deficient 2F-PSPCz containing two fluorine substituents, the energy of this state is unresponsive when changing from 2F-PSPCz to the even more electron-deficient 4F-PSPCz.

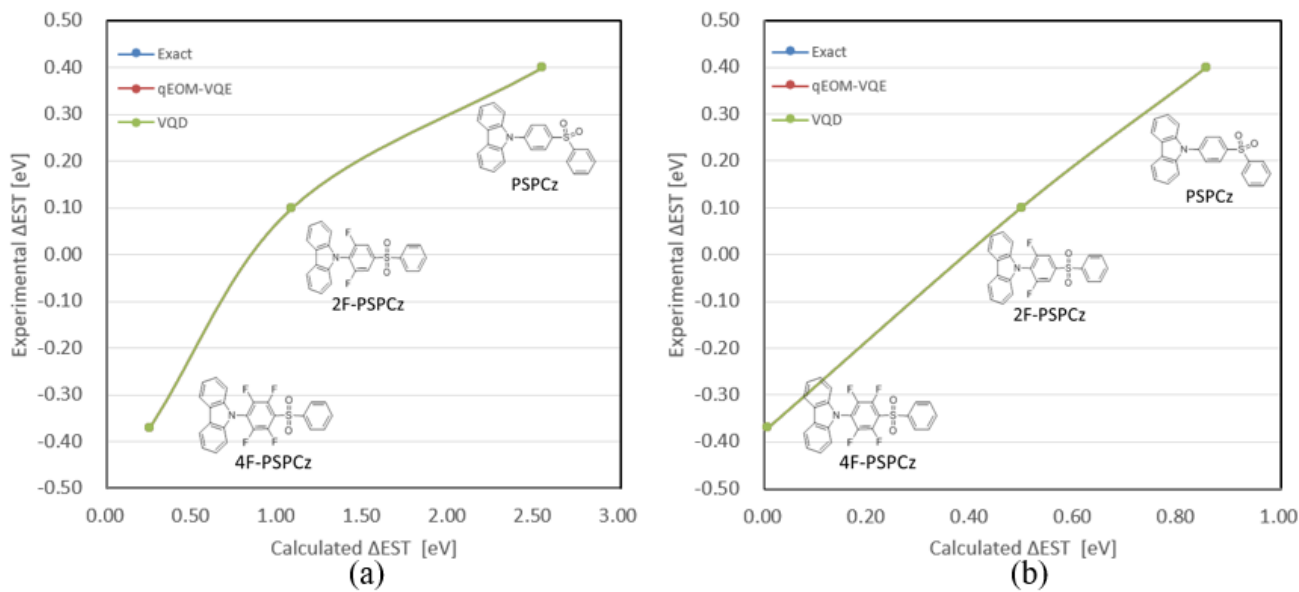

**Fig 5**. Correlations between $\Delta E_{st}$ from spectra experiments and exact diagonalization, qEOM-VQE and VQD calculations on state vector simulator using the (a) STO-3G and (b) 6-31G(d) basis sets.

In contrast, very different behavior was observed for simulations involving the 6-31G(d) basis set. The energies of both the $S_1$ and $T_1$ states are lowered by changing from PSPCz to 2F-PSPCz, while the energy of the $T_1$ state is raised for 4F-PSPCz in comparison to 2F-PSPCz. Notably, the $\Delta E_{st}$ gaps calculated by using the STO-3G basis set for computations on PSPCz, 2F-PSPCz and 4F-PSPCz are predicted to be 2.54 eV (93 *mHa*), 1.08 eV (41 *mHa*) and 0.25 eV (9 *mHa*), respectively. These are much larger than relative energies predicted by using the 6-31G(d) basis set which are 0.86 eV (31 *mHa*), 0.50 eV (18 *mHa*) and 0.0055 eV (0.2 *mHa*), respectively.

Direct comparison between laboratory experiments and simulations can be accomplished by comparing the $\Delta Est$ obtained from fluorescence spectra measured at room temperature and phosphorescence spectra measured at 77 K [18] with values predicted by Full CI and with *Ry* used with the qEOM-VQE and VQD algorithms. As shown in Figure 5, although $\Delta E_{st}$ gaps predicted by calculations with STO-3G provide good correlation with values obtained from experiments, the predictions are approximately an order of magnitude larger than those found in experiment. Much better agreement can be observed between values calculated with 6-31G(d) and experimental data, with calculations predicting energies that are approximately 0.4 eV (15 *mHa*) larger than experimental values. Note that the experimental value of $\Delta E_{st}$ for 4F-PSPCz is -0.4 eV (-15 *mHa*); this could be due to the fact that the energies of $S_1$ and $T_1$ were experimentally determined at different temperatures.

Overall, these results demonstrate that quantum computations performed with the qEOM-VQE or VQD algorithms utilizing the *Ry* Ansätze and the 6-31G(d) basis set on the HOMO-LUMO active space is an adequate method for predicting the $\Delta E_{st}$ of the PSPCz molecules of interest.

**B. Calculations with the `qasm` simulator**

Although we have shown that the energies of $S_1$ and $T_1$ can be accurately predicted by qEOM-VQE

and VQD methods on the `statevector` simulator, it is important to note that it is notoriously difficult to obtain similar accuracy by performing calculations on quantum devices due to the influence of device noise.

There are two major sources of noise that could affect the accuracies observed in this study. One is related to qubit quality and includes errors caused by depolarization, dephasing, and errors due to readout of qubits and deviations of quantum gates from their normal unitary operations. The other is related to sampling error caused by limits placed on measurements of single circuit executions (i.e. shots). The maximum number of shots possible with IBM Quantum devices and with the `qasm` simulator is 8192. We have evaluated whether this limit adequately samples the quantum circuits to accurately predict excited states of the TADF molecules.

qEOM-VQE and VQD calculations using the `qasm` simulator with 8192 shots were performed on PSPCz, 2F-PSPCz and 4F-PSPCz. Figure 6 shows plots of energies arising from these calculations. The maximum error with respect to the exact values for PSPCz, 2F-PSPCz and 4F-PSPCz were 0.6, 0.4 and 0.2 *mHa* from qEOM-VQE calculations and 1.0, 0.8 and 1.3 *mHa* from VQD calculations, respectively. Thus, excited state energies from both qEOM-VQE and VQD are within chemical accuracy (i.e. 1.6 *mHa*) of the energies obtained by exact diagonalization.

This agreement indicates that calculations derived from qEOM-VQE and VQD with 8192 shots accurately reproduce exact values of the $S_1$ and $T_1$ energies of tested TADF emitters with $\Delta E_{st}$ ranging from 0.0055 eV (0.2 *mHa*) to 0.86 eV (31 *mHa*).

The $\Delta E_{st}$ gaps computed for the PSPCz molecules by qEOM-VQE and VQD are compared with experimental results in Figure 7. The square of the correlation coefficient between $\Delta E_{st}$ gaps predicted by the use of both the qEOM-VQE and VQD algorithms and those obtained experimentally

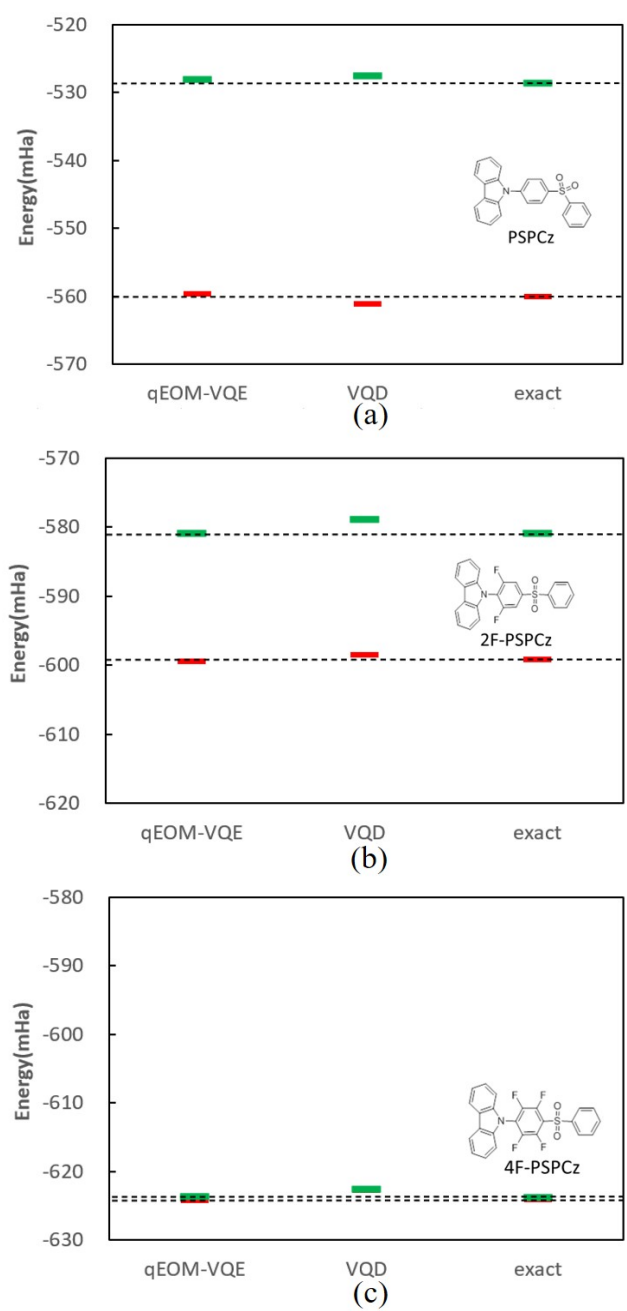

**Fig 6**. Energies of the $S_0$, $T_1$ and $S_1$ excited states (represented in blue, red and green) for (a) PSPCz, (b) 2F-PSPCz and (c) 4F-PSPCz from exact diagonalization, qEOM and VQD calculations on the `qasm` simulator with 8192 shots using the 6-31G(d) basis set.

are 0.999, indicating that 8192 measurement shots can satisfactorily predict these gaps.

We also found that simulations using qEOM-VQE reproduce the $\Delta E_{st}$ values predicted by Full CI, whereas simulations that utilize the VQD algorithm slightly overestimate these gaps. These results more sensitive to the number of measurement shots suggest that VQD is, at least in the present case, than qEOM-VQE, which may be due to the fact that VQD obtains excited state energies from a penalty term comprising the overlap between two quantum states. This overlap could possibly be improved by increasing the number of shots by a factor of 3 to 4; however, such improvements significantly increase the computational expense of quantum simulations using the `qasm` simulator.

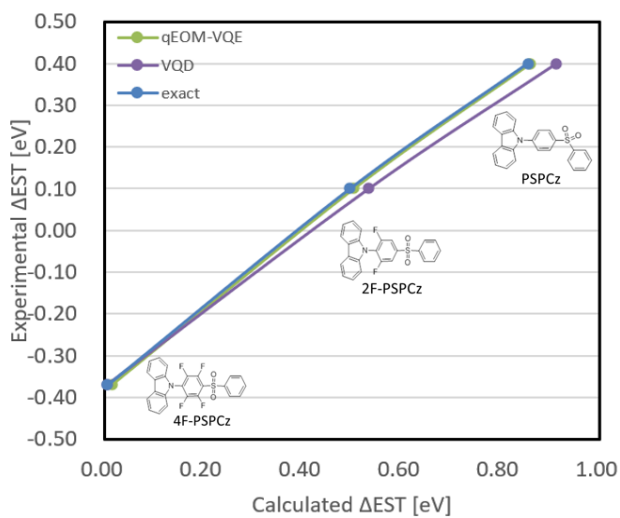

**Fig. 7**. Correlations between $\Delta E_{st}$ from spectral experiments and exact diagonalization, qEOM-VQE and VQD calculations on the `qasm` simulator with 8192 shots using the 6-31G(d) basis set.

**C. qEOM-VQE calculations on `ibmq_boeblingen`**

An adequate error mitigation approach must be applied to obtain an accurate ground state to use as a reference state in order to accurately compute excited state energies on a quantum device with the qEOM-VQE and VQD algorithms. Here we have applied the readout error mitigation and quantum state tomography techniques to ground states computed with VQE. This was followed by an examination of energies and the optimized $\theta$ parameters of the *Ry* Ansätze to determine whether results yielded by mitigation approaches are in good agreement with those generated by Full CI.

The energies of the $S_0$ ground states of the PSPCz, 2F-PSPCz and 4F-PSPCz molecules computed with the *Ry* Ansätze using the VQE

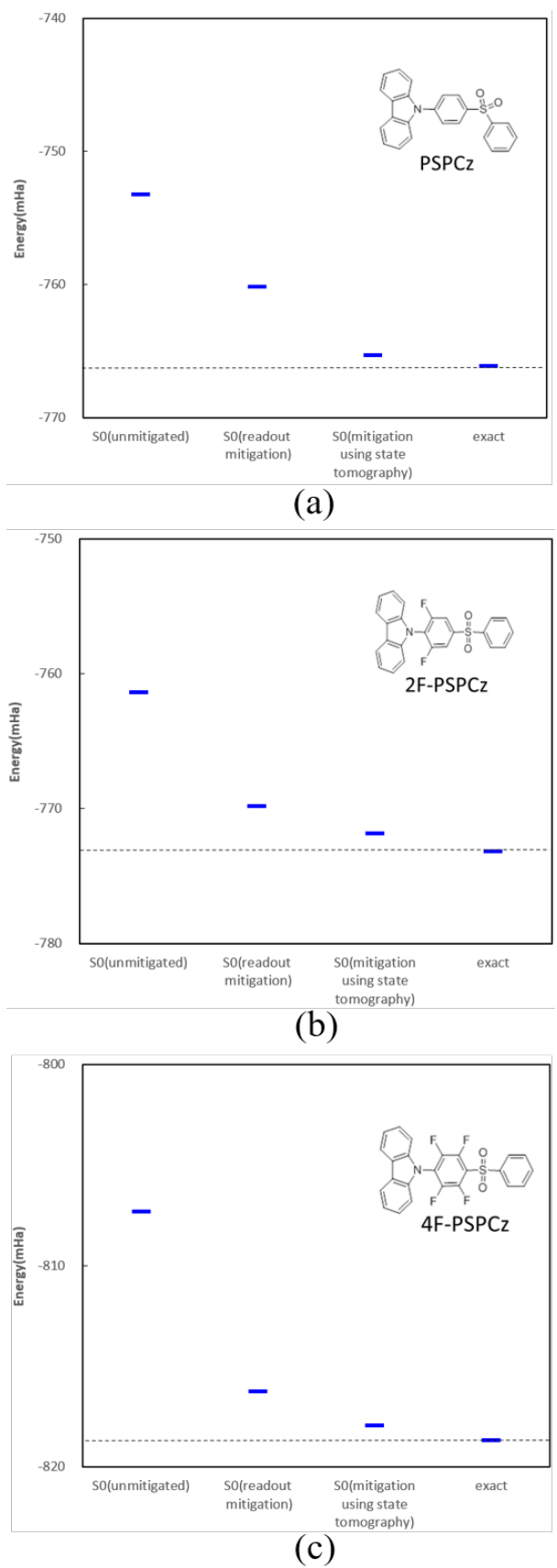

**Fig 8**. Electronic energies, in *mHa*, of the $S_0$ state from VQE calculations on the `ibmq_boeblingen` device and from Full CI for (a) PSPCz, (b) 2F-PSPCz and (c) 4F-PSPCz.

algorithm on the `ibmq_boeblingen` device with and without error mitigation approaches are shown in Figure 8. VQE calculations without error mitigation yields energies about 13 *mHa* higher than exact values. This difference is improved to about 6 *mHa* by the use of readout error mitigation. The use of quantum state tomography yields energies that are within 1 *mHa* of the exact values and is, therefore an even more accurate approach than readout error mitigation.

These results suggest that errors generated by performing noisy VQE calculations arises from a combination of readout error and decoherence noise. While readout error can be reduced by application of either readout error mitigation and quantum state tomography, the decoherence noise (which is mainly due to depolarization) can only be corrected by quantum state tomography.

To examine the difference between the ground state obtained from Full CI and the state obtained from the final parameter-optimized *Ry* Ansätze from quantum device, we have compared energies obtained by the use of Full CI with those obtained from the use of `statevector` simulations with the parameter-optimized *Ry* Ansätze obtained from quantum state tomography. Figure 9 and Table 1 (Appendix II) show that VQE energies obtained from calculations with the quantum state tomography-mitigated parameter-optimized *Ry* Ansätze always lie within 1 *mHa* of ground states obtained with Full CI, suggesting that the final states obtained by the use of the quantum state tomography approach are similar to the ground states obtained with Full CI.

We unexpectedly found that energies obtained by the use of readout error mitigation on results obtained from the *Ry* Ansätze using the VQE algorithm on the `ibmq_boeblingen` device are almost similar to, or slightly higher than, energies obtained from calculations without readout error mitigation. These results indicate that the optimized *R*y Ansätze obtained from readout error mitigation approach are no closer to the ground state than the optimized *R*y Ansätze obtained without error mitigation approaches. Evidently, readout error mitigation fails to produce the pure state of the ground state even after optimization of the parameters of the *Ry* Ansätze.

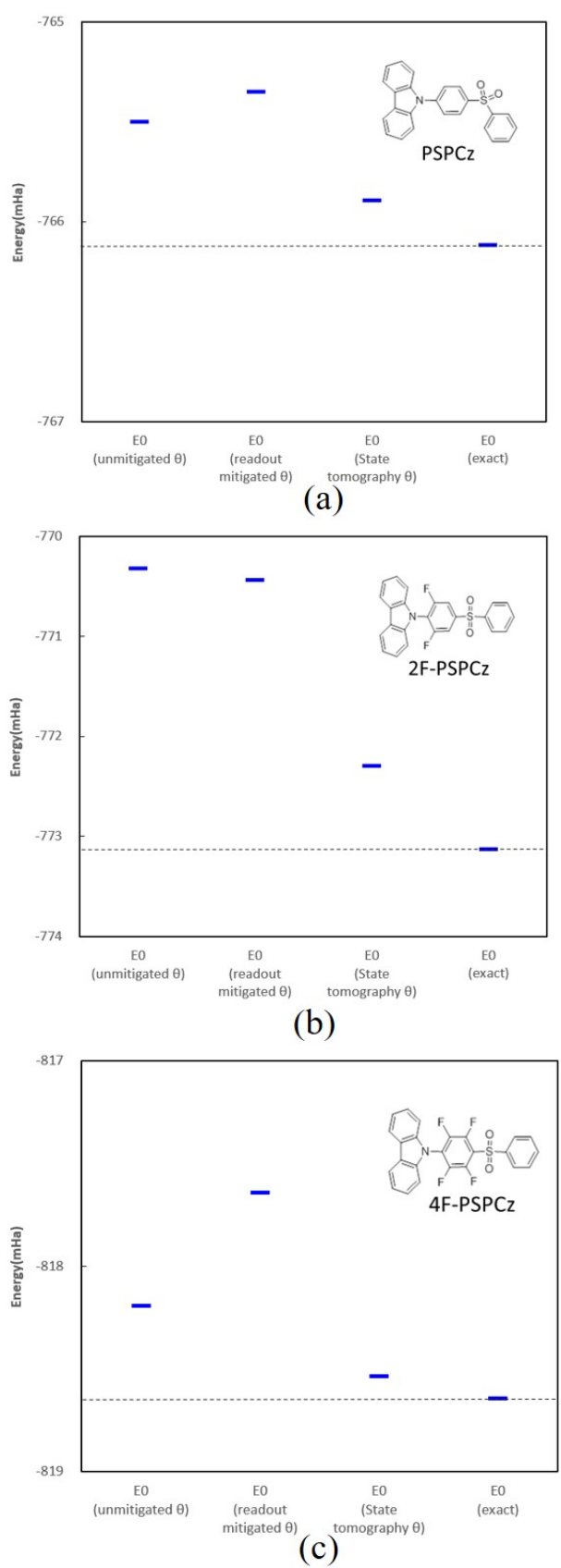

**Fig 9**. Electronic energies of the $S_0$ ground states of PSPCz molecules obtained from `statevector` simulations with *Ry* and Full CI. Optimized parameters for *Ry* were determined from VQE calculations on `ibmq_boeblingen`.

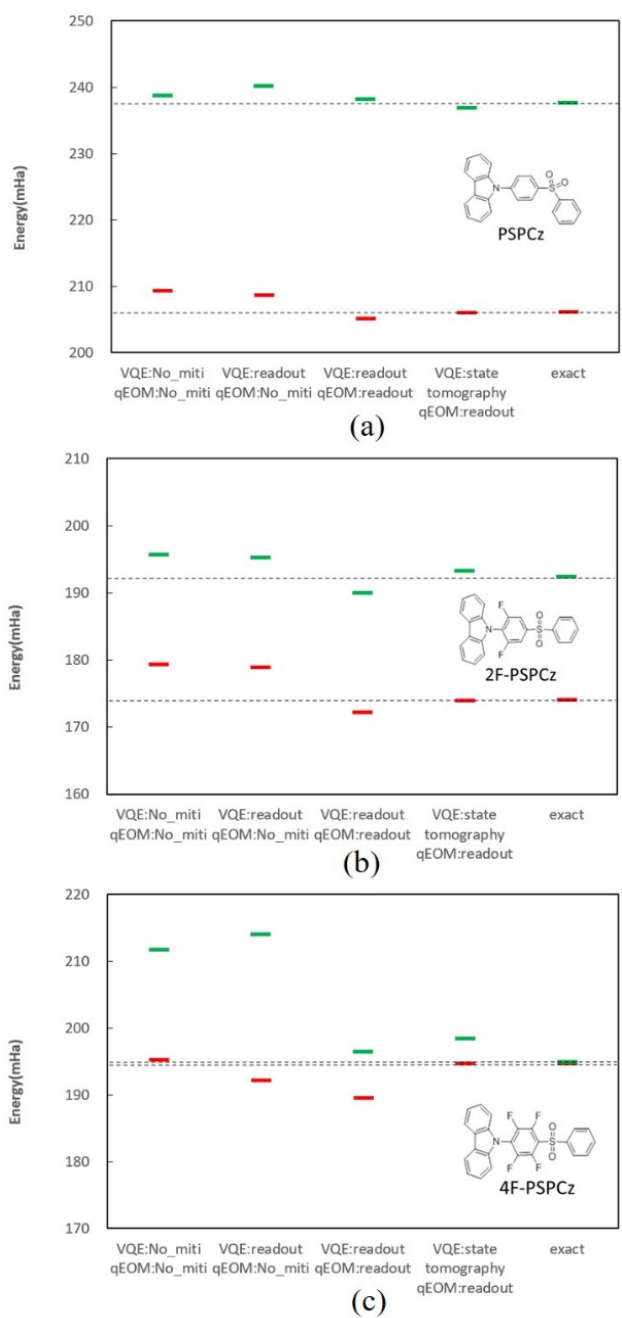

**Fig 10.** Relative energies of the $T_1$ and $S_1$ excited states (shown in red and green, respectively) of PSPCz, 2F-PSPCz and 4F-PSPCz with respect to the $S_0$ ground state obtained from qEOM-VQE calculations on `ibmq_boeblingen` and from Full CI.

We undertook an examination of the effect of error mitigation approaches on the accuracy of excited state energies predicted by the use of the *Ry* Ansätze with the qEOM-VQE algorithm for the carbazoles of interest to this study. Figure 10 and Table I (Appendix II) show that energies of excited states obtained from qEOM-VQE/*Ry* without error mitigation approaches are 3, 5 and 16 *mHa* larger than energies obtained by Full CI for the PSPCz, 2F-PSPCz and 4F-PSPCZ molecules, respectively. These results indicate that the qEOM-VQE algorithm overestimates the energies of $S_1$ or $T_1$ states without error mitigation being applied, particularly if both states are energetically close.

When readout error mitigation is applied to the ground state computed with VQE alone and not to measurements of the matrix elements provided by the EOM, only marginal improvement is observed for the error of the energy computed for 2F-PSPCz, but the errors increase for calculations involving PSPCz and 4F-PSPCZ. In contrast, applying readout error mitigation to *both* the energy of ground state computed with VQE/*Ry* and to measurements of the matrix elements provided by EOM improves the accuracy of excited state energies compared to calculations without readout error mitigation.

Notably, the largest observed difference between the excited state energy computed by the qEOM-VQE algorithm and by Full CI was 16 *mHa*, found for calculations involving 4F-PSPCZ, but this value improves to 1 *mHa* after inclusion of readout error mitigation to both the VQE and qEOM-VQE calculations involving the ground and excited states, respectively. These results suggest that failure to accurately compute values of the matrix elements of the EOM is one of the main factors that contribute to the deviation of qEOM-VQE excited state energies from those predicted by Full CI.

The best predictions of excited state energies were found by applying quantum state tomography techniques to obtain an accurate ground state from VQE calculations, and then applying this reference state to the readout error mitigated measurement of the matrix elements of the EOM for the excited states obtained from the qEOM-VQE algorithm. This procedure results in energies that are within 3 *mHa* of those obtained by the use of Full CI.

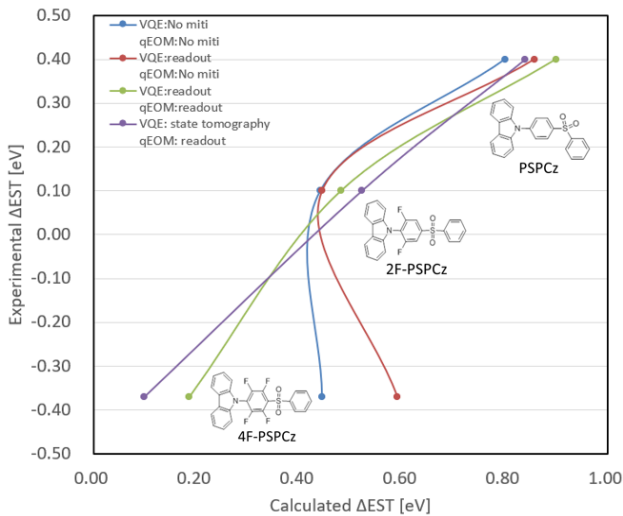

**Fig 11**. Correlations between $\Delta E_{st}$ for the PSPCz molecules from spectral experiments and qEOM-VQE calculations with the *Ry* Ansätze on ibmq_boeblingen using the 6-31G(d) basis set.

Next we compared the $\Delta E_{st}$ of the PSPCz molecules obtained from qEOM-VQE calculations performed on `ibmq_boeblingen` with results obtained from laboratory experiments (Figure 11 and Table I; Appendix II). For calculations performed without error mitigation approaches, good agreement was observed between experiment and computational predictions involving PSPCz and 2F-PSPCz, but a larger value of $\Delta E_{st}$ was predicted for 4F-PSPCz than was found in experiment.

Unfortunately, the use of readout error mitigation for the VQE calculation of the ground state did not result in improvement of the predicted $\Delta E_{st}$ of any of the phenylcarbazole molecules. However, a much better prediction of the $\Delta E_{st}$ was obtained by applying readout error mitigation to the measurements of the matrix elements of the EOM.

This conclusion is further bolstered by the fact that the application of quantum state tomography techniques to ground states predicted by VQE resulted in excellent agreement between computational predictions of $\Delta E_{st}$ and experiments for all the examined PSPCz molecules. Overall, these results indicate that the best results are obtained by applying quantum state tomography techniques to the ground state computed with VQE and applying readout error mitigation to the measurements of the matrix elements of the EOM for qEOM-VQE computations of excited states.

**D. VQD calculations on `ibmq_singapore`**

Using VQD method for calculating excited states of tested TADF emitters on `statevector` and `qasm` simulators shows great promise, but in principle there is a fundamental problem associated with performing VQD calculations on quantum device that will need to be addressed having to do with the fact that device noise influences the overlap term, which is necessary to search the excited states, and the molecular Hamiltonian term used to calculate the energy of excited states. In this section, we will describe how the overlap and molecular Hamiltonian terms of TADF molecules are affected by the device noise, which consequently significantly limit the accuracy of VQE calculation results.

Figure 12 and Table I (Appendix II) compares excited state energies of 2F-PSPCz using the same VQD calculation protocol but with two different quantum reference states to calculate the overlap term: (a) those obtained by purifying VQD results with quantum state tomography (b) and those left unpurified. The values of the overlap term for the first excited state, $T_1$, with unpurified and purified reference states converges to similar values, ~0.04, indicating that VQD satisfactorily computes the excited state. On the other hand, the excited state energy obtained by the calculation using the unpurified reference state is 20 *mHa* higher than the exact value, in contrast to that obtained by using the purified reference state which is only 8 *mHa* higher. These results indicate that the excited state computed by using the purified reference state possesses more fidelity with the ground state than that obtained by using the unpurified reference state.

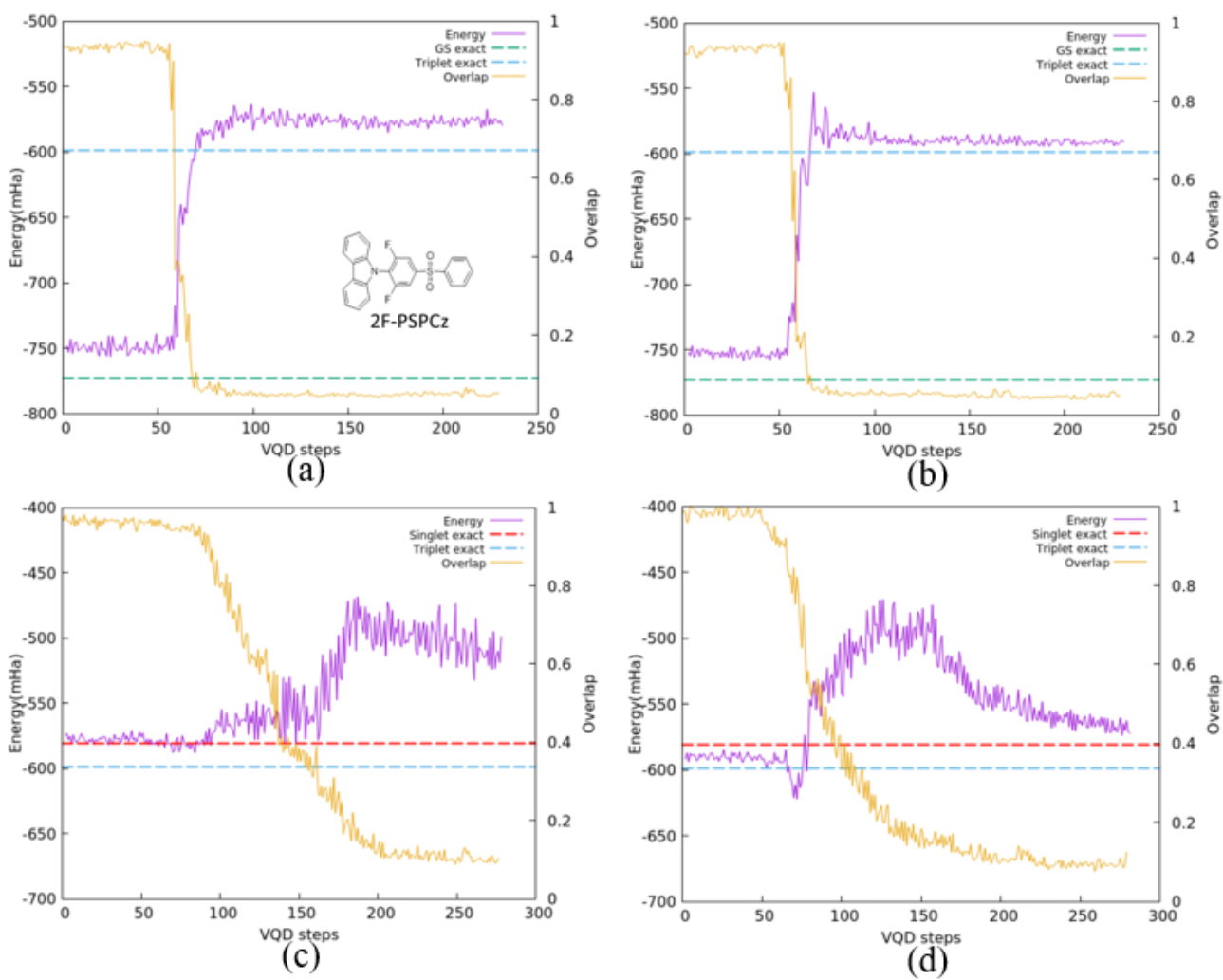

**Fig 12**. VQD iterations for calculations of 2F-PSPCz on the `ibmq_singapore` quantum devices with the SPSA optimizer for (a) $T_1$ using the unpurified reference state, (b) $T_1$ using the reference state purified by quantum state tomography, (c) $S_1$ using the unpurified reference state (d) $S_1$ using the reference state purified by quantum state tomography.

More importantly, the use of the unpurified reference state leads to energies of the second excited state, $S_1$, that can be as large as 88 *mHa* higher than the exact energies obtained with FCI, whereas the use of the purified reference state results in $S_1$ energies that are only 12 *mHa* larger than the exact energies. These results show that a large overlap term can lead to large errors for calculations involving quantum states that have not been purified by quantum state tomography.

Figure 12 also indicates that the accuracy of VQD calculations may be invariant to changes in the level of excited state when quantum states purified by quantum tomography are used, which supports our contention that quantum states purified by quantum state tomography are very close to the exact state.

Next, we used the purified reference state to investigate excited state energies estimated by the molecular Hamiltonians of PSPCz, 2F-PSPCz and 4F-PSPCz as shown in Figure 13 and Table 1. Without error mitigation, device noise causes the energies of computed excited states to deviate from exact values by 8, 12 and 15 *mHa* for PSPCz, 2F-PSPCz and 4F-PSPCz, respectively. Error mitigation using state tomography improves these deviations to 0, 2 and 4 *mHa* for PSPCz, 2F-PSPCz and 4F-PSPCz, respectively, signifying the effectiveness of this technique to correct errors obtained by using VQD.

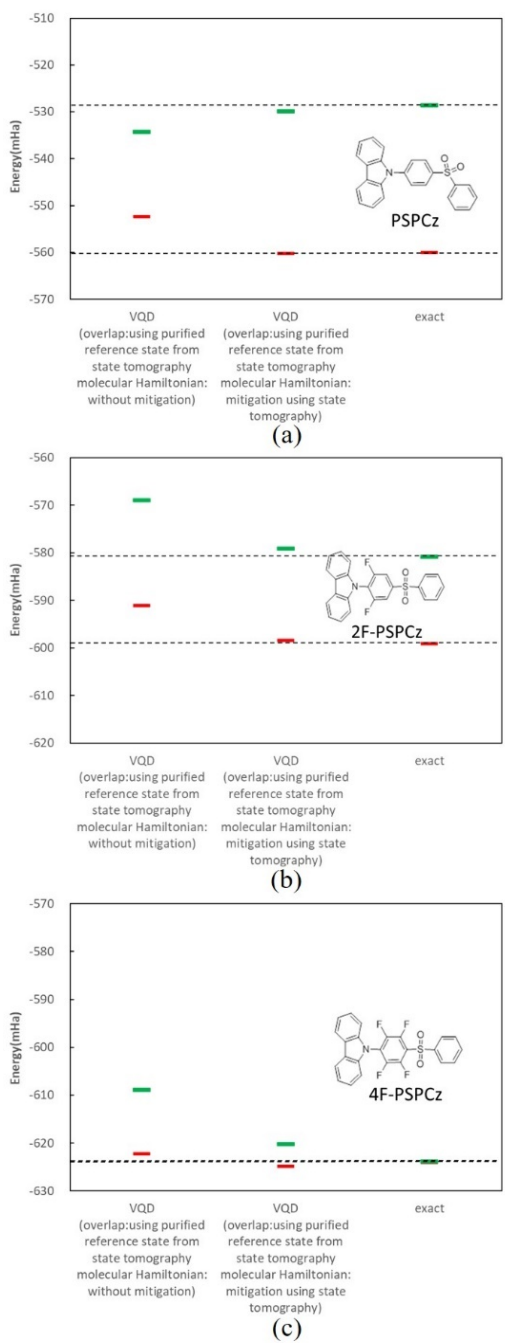

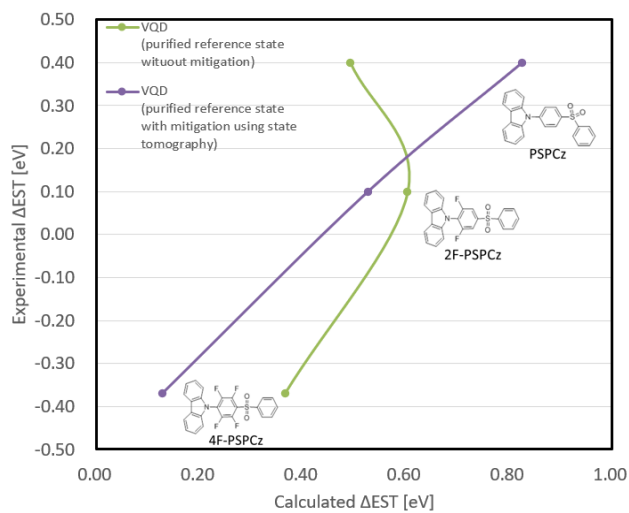

**Fig 13.** Energies of the T$_1$ and S$_1$ excited states (shown in red and green, respectively) of PSPCz, 2F-PSPCz and 4F-PSPCz obtained from VQD calculations on `ibmq_singapore` and from Full CI.

**Fig 14.** Correlations between $\Delta E_{st}$ from spectral experiments and VQD calculations with the *Ry* Ansätze on `ibmq_singapore` using the 6-31G(d) basis set.

Figure 14 shows comparisons of $\Delta E_{st}$ from experimental data with results computed with VQD. Those results demonstrate that unmitigated results deviate from experiment. For example, the $\Delta Est$ of experimental data follows the trend: PSPCz > 2F-PSPCz > 4F-PSPCz, whereas the computed $\Delta E_{st}$ exhibits the following order: 2F-PSPCz > PSPCz > 4F-PSPCz. On the other hand, when state tomography is used, $\Delta E_{st}$ were found to be in good agreement with the experimental data. These results suggest that VQD calculations that utilize the state tomography technique reliably predicts $\Delta E_{st}$ of TADF emitters. Overall, the state tomography technique does an excellent job in calculating both overlap and molecular Hamiltonian terms on quantum devices in order to reproduce exact energies of excited states to an excellent degree, and reliably predict experimental values of $\Delta E_{st}$ gaps.

**Conclusion**

Excited states of three phenylsulfonyl-carbazole TADF emitters used in OLED display were investigated by qEOM-VQE and VQD calculations using an active space of HOMO and LUMO on the statevector and qasm simulators and IBM Quantum devices. Calculations performed with the qEOM-VQE and VQD algorithms on statevector and qasm simulators accurately reproduced the results from exact diagonalization and provided reliable prediction of $\Delta E_{st}$ which are in good agreement with experiments. These results show that the qEOM-VQE and VQD algorithms are accurate and can be reliably used to develop a fundamental understanding of the relation between structural variation and energies of excited states that could aid in the design of novel TADF emitters.

Results from qEOM-VQE and VQD calculations on IBM Quantum devices show that noise significantly limits the accuracy of predicted values of excited states. The excited energies were predicted to differ from exact values by as much as 16 *mHa* and 88 *mHa* by simulations performed with the qEOM-VQE and VQD algorithms, respectively. A new practical scheme which utilizes quantum state tomography was used to calculate excited states on quantum devices. By using this scheme, excited state energies from both qEOM-VQE and VQD calculations could be computed that are within 3 *mHa* of energies obtained by Full CI and the corresponding singlet-triplet gaps, $\Delta E_{st}$, are in good agreement with experimental data. We note that our work is the first validated study of excited states using the VQD method which is an extension of VQE. This work has highlighted the utility of evaluating excited states with state tomography on quantum devices.

**Appendix I**

Detailed procedure for the practical application of quantum state tomography for VQD simulations on an IBM Quantum device:

(i) Compute the ground state $|\Psi_0\rangle$ by VQE to minimize the cost function

$$C_0(\theta) = \langle\Psi(\theta)| H |\Psi(\theta)\rangle$$

with respect to the parametrized Ansätze state

$$|\Psi(\theta)\rangle = U(\theta) |0\rangle$$

where $U(\theta)$ is a unitary transformation with parameter $\theta$. Ideally, this procedure produces the purified ground state

$$|\Psi_0\rangle = |\Psi(\theta_*)\rangle = U(\theta_*) |0\rangle$$

In practice, the readout error mitigation technique can be used to compute $C_0(\theta)$ and thereby find $\theta_*$.

(ii) The ground state obtained in step (i) is then used to find the first excited state with VQD by minimizing the following modified cost function:

$$\langle\Psi(\theta)| H |\Psi(\theta)\rangle + \beta_0|\langle\Psi(\theta)|\Psi_0\rangle|^2$$

with a suitably chosen weighting parameter, $\beta_0$ as described in ref. [6]. The second term denotes a state that minimizes the energy under the constraint that $|\Psi(\theta)\rangle$ is orthogonal to the reference state $|\Psi(\theta_*)\rangle$. Ideally, the minimizer of $C_1(\theta)$ is the 1st excited state. On the other hand, the final state of a quantum operation $U(\theta_*)$ implemented on current noisy quantum devices, must be a mixed state $\rho_0$, rather than the pure state $|\Psi(\theta_*)\rangle$, and thus the following alternative cost function can be considered:

$$\langle\Psi(\theta)| H |\Psi(\theta)\rangle + \beta_1\langle\Psi(\theta)|\rho_0|\Psi(\theta)\rangle$$

where $\rho_0$ is found via state tomography (without readout error mitigation). Note that instead of $\rho_0$, we have defined the cost function in this manuscript by using the largest eigenvector of $\rho_0$, denoted by $|\Psi'_0\rangle$, such that the cost function is given by

$$C'_1(\theta) = \langle\Psi(\theta)| H |\Psi(\theta)\rangle + \beta_0|\langle\Psi(\theta)|\Psi'_0\rangle|^2$$

The (approximate) first excited state is then given by the minimized $C'_1(\theta)$. The procedure adopted for the investigations described in this manuscript have compared the minimizers of $C_1(\theta)$ and $C'_1(\theta)$.

(iii) The second excited state is computed on a quantum device by utilizing a process similar to that used for the ground state. First, the final state of the mixed first excited state that minimizes $C'_1(\theta)$ is purified by constructing the density matrix via state tomography. The largest eigenvector of this matrix, denoted by $|\Psi'_1\rangle$, is used to minimize the cost function under the constraint that $|\Psi(\theta)\rangle$ is orthogonal to two reference states $|\Psi'_0\rangle$ and $|\Psi'_1\rangle$.

$$C'_2(\theta) = \langle\Psi(\theta)| H |\Psi(\theta)\rangle + \beta_0|\langle\Psi(\theta)|\Psi'_0\rangle|^2 \\ + \beta_1|\langle\Psi(\theta)|\Psi'_1\rangle|^2$$

This procedure can then be repeated to compute higher excited states.

**Appendix II**

**Table 1.** Energy deviations, in *mHa*, for simulations performed on a quantum device in comparison to FCI

**Ground state energy deviations for VQE simulations in comparison to FCI**

| *TADF emitter* | *unmitigated* | *readout mitigated* | *state tomography mitigated* |
|---|---|---|---|
| PSPCz | 13 | 6 | 1 |
| 2F-PSPCz | 12 | 3 | 1 |
| 4F-PSPCz | 11 | 2 | 1 |

**Excited state energy deviations for qEOM-VQE simulations in comparison to FCI**

| *TADF emitter* | *State* | *VQE: unmitigated/ qEOM: unmitigated* | *VQE: readout mitigated/ qEOM: unmitigated* | *VQE: readout mitigated/ qEOM: readout mitigated* | *VQE: state tomography mitigated/ qEOM: readout mitigated* |
|---|---|---|---|---|---|
| PSPCz | $T_1$ | 3 | 3 | -1 | 0 |
| | $S_1$ | 1 | 3 | 1 | -1 |
| 2F-PSPCz | $T_1$ | 5 | 5 | -2 | 0 |
| | $S_1$ | 3 | 3 | -2 | 1 |
| 4F-PSPCz | $T_1$ | 1 | -2 | -5 | 0 |
| | $S_1$ | 16 | 19 | 1 | 3 |

**Excited state energy deviations for VQD simulations in comparison to FCI**

| *TADF emitter* | *State* | *overlap using the unpurified reference state/ unmitigated molecular Hamiltonian* | *overlap using purified reference state from state tomography/ unmitigated molecular Hamiltonian* | *overlap using purified reference state from state tomography/ state tomography mitigated molecular Hamiltonian* |
|---|---|---|---|---|
| PSPCz | $T_1$ | 21 | 8 | 0 |
| | $S_1$ | -11 | -6 | 1 |
| 2F-PSPCz | $T_1$ | 23 | 8 | 1 |
| | $S_1$ | 88 | 12 | 2 |
| 4F-PSPCz | $T_1$ | 79 | 2 | -1 |
| | $S_1$ | 82 | 15 | 3 |